# Large Longitudinal Magnetoelectric Coupling in $NiFe_2O_4$-$BaTiO_3$ Laminates

**Deepak Patil, June-Hee Kim,[1] Yi Sheng Chai, Joong-Hee Nam,[1] Jeong-Ho Cho,[1] Byung-Ik Kim,[1] and Kee Hoon Kim[*]**

*CeNSCMR, Department of Physics and Astronomy, Seoul National University, Seoul 151-747, Republic of Korea*

[1]*Center for Electronic Component Research,, Korea Institute of Ceramic Engineering and Technology, Seoul 153-801, Republic of Korea*

## Abstract

In contrast to the Pb-based magnetoelectric laminates (MELs), we find in the $BaTiO_3$ and $NiFe_2O_4$ laminates (number of layers $n$=5-25) that the longitudinal magnetoelectric (ME) voltage coefficient $\alpha_{E33}$ becomes much larger than the transverse one due to preferential alignment of magnetic moments along the $NiFe_2O_4$ plane. Moreover, upon decreasing each layer thickness down to 15 μm, we realize enhanced $\alpha_{E33}$ up to 18 mV/cmOe and systematic increase of the ME sensitivity in proportion to $n$ to achieve the largest in the Pb-free MELs (400×10$^{-6}$ V/Oe), thereby providing pathways for tailoring ME coupling in mass-produced, environment friendly laminates.

[*] Email address: khkim@phya.snu.ac.kr

Magnetoelectric (ME) composites consisting of magnetostrictive and piezoelectric materials[1-4] offer a unique capability of cross-control of magnetization (*M*) and polarization (*P*) by an electric field (*E*) and a magnetic field (*H*), which can be potentially useful in, for example, an energy harvester[5-7] or a field sensor.[8,9] Especially, they show relative higher power density than the conventional methods for vibration energy harvesting.[10-13] In such ME composites, *P* is varied by *H* via the strain coupling at the interface between the two phases.[14,15] Early works were primarily focused on the ME composites made of randomly mixed particulates of each component.[16] However, to effectively increase the interface coupling strength, magnetoelectric laminates with alternative magnetostrictive and piezoelectric thick films have been fabricated and showed much better ME properties.[17,19]

Most of these laminates employed $Pb(Zr,Ti)O_3$ or $Pb(Mg_{1/3}Nb_{2/3})O_3$-$PbTiO_3$ as a piezoelectric material. Although quite a large ME voltage coefficient $α_E = δE/δH$ was achieved (>1V/cmOe),[20] those laminates are indeed susceptible to pollution and toxicity due to Pb.[21] It is also noteworthy that in most of the known laminates, $α_E$ was much larger in a transverse configuration (i.e., $E \perp H$) rather than in a longitudinal one (i.e., $E//H$). The effect essentially originates from the fact that $α_E$ in the laminate structure is determined by the in-plane magnetostriction, which can be tuned effectively by the in-plane magnetic field.[4,16] On the other hand, as the Pb-free piezoelectric ceramics are in increasing demand in various industrial products, it is becoming increasingly important to develop Pb-free magnetoelectric laminates (MELs) as well. Furthermore, it is equally important to develop a fabrication process of mass producing MELs with low cost. Recently, Israel et al.[22)] have indeed found ME properties in a cheap (one cent valued), mass-produced, commercial magnetoelectric laminates that has 81 layers of $BaTiO_3$ film sandwiched by ferromagnetic Ni electrodes; they found the maximum ME voltage sensitivity $\delta V/\delta H = 7 \times 10^{-6}$ V/Oe and $\alpha_E$ =7 mV/cmOe. In this commercial MELs targeted to enhance the capacitance, however, the Ni electrodes are connected in parallel (left in Fig. 1(a)) so that it is optimal to detect the modulation of charges, not voltage, under $H$ bias. Thus, development of the series-connected MELs (right in Fig. 1(a)) is a promising approach to induce significantly enhanced ME voltage since the voltage can be increased in proportion to the number of layers. Moreover, the ME properties of the series-connected MELs are expected to be further tailored by for example, selection of proper constituent materials with higher piezoelectricity/magnetostriction or optimizing strain coupling at the interface. In this study, we report the successful synthesis and characterization of the Pb-free MELs, composed of alternating $n$ ($n$=5-25) layers of $BaTiO_3$ (BTO) and $NiFe_2O_4$ (NFO) thick films. In contrast to most of the previously known MELs, we find that the particular MEL exhibits five times larger longitudinal $\alpha_E$ than the transverse one, due to the predominant alignment of the magnetic moment into the laminate plane.

The MEL structure was synthesized by the tape casting method[23)] to form alternating, thick layers of NFO and BTO with the same layer thickness and a typical lateral dimension of ~2×2 $mm^2$. The number of BTO layers, $n$, was varied from 5 to 25 with the thickness $t$ of every layer varied from 15 to 50 μm. After polishing the surface of the MEL, electrical contacts were made on the outer NFO layers using silver paste and samples were poled at room temperature by applying a dc electric field of 100 kV/cm. An ac ME voltage $\delta V$ across the sample, in response to a driven ac field $\delta H$ = 3.36 Oe at a low frequency of 194 Hz, was measured by a lock-in amplifier as a function of dc magnetic field, $H_{dc}$. The $\delta V$ measurements along the out-of-plane direction were performed for both in-plane and out-of-plane $H_{dc}$ (and $\delta H$) orientations to estimate the transverse ($\alpha_{E31}$) and longitudinal ($\alpha_{E33}$) ME voltage coefficients, respectively (See also, Fig. 1(b)). Magnetostriction ($\lambda$) was measured by a strain gauge (Kyowa, KFL-02-120-C1-11) while magnetization was measured by a vibrating sample magnetometer.

Figure 1(b) displays a typical optical microscopy image of the MEL sample with a layer thickness $t$ = 50 μm. The alternating BTO and NFO layers, which are clearly distinguished by well defined interfaces, are formed homogeneously over the vertical dimension of the MEL. We also measured the X-ray diffraction (XRD) of the MEL both before and after grinding to a powder form. As shown in Fig. 1(c), both XRD patterns reveal pure NFO and BTO phases without any impurity, showing that each phase is retained in the MEL. However, we note that the peak positions of BTO and NFO in the MEL are located at larger angles than those in the powder forms (inset of Fig. 1(c)). This implies the existence of compressive strain in the MEL structure, possibly induced during the synthesis.

Shown in Fig. 2(a) is the $H_{dc}$ dependence of $\alpha_{E33}$ and $\alpha_{E31}$ for the MELs with $n = 15$ and $t = 50$ μm. $\alpha_{E33}(H)$ clearly shows a sign reversal with reversing $H_{dc}$ and reaches its maximum near ±2 kOe, which are typical ME signals expected in strain-coupled media.[24] The $\alpha_{E31}(H)$ also shows a qualitatively similar behavior. On the other hand, $\alpha_{E31}(H)$ shows noticeable hysteresis and has 5 times smaller peak magnitude than $\alpha_{E33}(H)$. It is noted that a similar magnitude difference between $\alpha_{E31}(H)$ and $\alpha_{E33}(H)$ was also observed in the MELs with different $n$ values.

It has been known that $\alpha_{E31}$ and $\alpha_{E33}$ in laminated ME media can be described through the following equations:

$$\alpha_{E33} = \frac{2kf(1-f)d_{31}\mu_0 \underline{s}}{(2fd_{31}^2 - \varepsilon_{33}\underline{s})[\overline{\mu}\underline{s} + 2kq_{13}^2(1-f)^2]} q_{13}, \tag{1}$$

$$\alpha_{E31} = \frac{-kf(1-f)d_{31}(q_{11}+q_{21})}{(\varepsilon_{33}\underline{s} - 2kfd_{31}^2)}, \tag{2}$$

where $\underline{s} = f(s_{11}^p + s_{12}^p) + k(1-f)(s_{11}^m + s_{12}^m)$ and $\overline{\mu} = f\mu_0 + (1-f)\mu_{33}^m$. Here, $f$ is the volume fraction of the magnetic phase, $k$ is the interface coupling parameter and $k = 1$ for an ideal interface and $k = 0$ for the case without frictions, $s_{ij}^{p(m)}$ are compliances for piezoelectric (magnetic) components, $d_{31}$ is the transverse piezoelectric coefficient, $q_{ij}$ are the piezomagnetic coefficients, i.e., $d\lambda_{ij}/dH_j$, where $\lambda_{ij}$ is the magnetostriction in the $i$-axis for $H$ along the $j$-axis, and $\mu_0$ and $\mu_{33}^m$ are the permeability of the free space and the magnetic phase, respectively. According to these equations, $\alpha_{E33}(H)$ and $\alpha_{E31}(H)$ should be roughly proportional to $q_{13}(H)$ and $q_{11}(H)+q_{21}(H)$, respectively. To confirm this, we measured the $H$-dependence of magnetostriction $\lambda_{ij}$ in the MEL with $n = 15$ and $t = 50$ μm (Fig. 2(b)) and calculated corresponding $q_{ij} = d\lambda_{ij}/dH_j$ in Fig. 2(c). We indeed find in Figs. 2(a) and 2(c) that the experimental $\alpha_{E33}(H)$ and $q_{13}(H)$ data have very similar curvatures and the same peak fields. On the other hand, we note that $\lambda_{11}(H) + \lambda_{21}(H) + \lambda_{31}(H) \approx 0$ due to the volume conservation of the sample under the application of the in-plane $H_1$ and thus the relationship $q_{11} + q_{21} \approx -q_{31}$ should hold. Then, eq. (2) predicts that $\alpha_{E31}(H) \propto -q_{31}(H)$. In the experimental data shown in Figs. 2(a) and 2(c), we can find that $\alpha_{E31}(H)$ and $-q_{31}(H)$ show similar hysteresis behavior and nearly the same peak magnetic fields. All the above results support the validity of eqs. (1) and (2) and suggest that the $H$-dependence of ME signals in the MELs can be mainly determined by that of the piezomagnetic coefficients of the magnetostrictive materials.

For more quantitative comparison on the magnitudes of $\alpha_{E33}(H)$ and $\alpha_{E31}(H)$, we can derive the following equation from eqs. (1) and (2) for the $k = 1$ case:

$$\frac{\alpha_{E33}}{\alpha_{E31}} = \frac{2q_{13}}{(q_{11}+q_{21})} \left( \frac{\mu_0 \underline{s}}{[\overline{\mu}\underline{s} + 2q_{13}^2(1-f)^2]} \right) \approx \frac{2q_{13}}{(q_{11}+q_{21})} \left( \frac{\mu_0}{\overline{\mu}} \right) , \qquad (3)$$

Since it is known that $\mu_{33}^m \approx 3\mu_0$ for NFO and $f \approx 0.5$ in the present MEL structure, we obtain $\overline{\mu} \approx 2\mu_0$ and thus $\alpha_{E33}/\alpha_{E31} \approx q_{13}/(q_{11}+q_{21}) \approx q_{13}/-q_{31}$. Direct comparison of Figs. 2(a) and 2(c) reveals that the ratio between the maximum values of $\alpha_{E33}$ and $\alpha_{E31}$ is 5.3, which is roughly close to that between $q_{31}$ and $-q_{13}$ (=4.3). This observation directly supports that the large $\alpha_{E33}/\alpha_{E31}$ value results from the large value of $q_{13}/-q_{31}$ in our MEL structure. More specifically, a large $\lambda_{13}$ (or $q_{13}$) i.e., large '1' axis change under $H_3$ as well as small $\lambda_{31}$ (or -$q_{31}$), i.e., minimal '3' axis change under $H_1$, can be mainly responsible for the unusually large $\alpha_{E33}/\alpha_{E31}$ = 5.3.

The observation of a much smaller $|\alpha_{E31}|$ than $|\alpha_{E33}|$ in our MEL is quite unusual. In most of the laminate structures studied so far, it has been reported that the transverse configuration produces larger ME signals.[4,25] To properly understand the physical origin of this enhanced $\alpha_{E33}$, we thus tried to determine the related magnetic anisotropy of the present MEL. Figure 2(d) shows that the magnetizations are saturated easily by the in-plane magnetic fields (i.e., $H_1$ and $H_2$) rather than the out-of-plane field (i.e., $H_3$), demonstrating the predominant in-plane alignment. Furthermore, there was no magnetic anisotropy in the plane so that the $M$ curves for $H_1$ and $H_2$ overlapped each other. Consistent with this, we could confirm that $\lambda_{21}(H) = \lambda_{12}(H)$ (not shown in the figures). All these features in the magnetization curves of our MEL directly suggest that actual magnetic moments mostly lie in the NFO plane and possibly demagnetized within the plane, as schematically drawn in the inset of Fig. 2(d). The predominant in-plane alignment of magnetic moments is likely to be associated with the internal compressive strain in the NFO layers as found in the XRD data (Fig. 1(c)). Moreover, in such a demagnetized domain pattern, an in-plane $H$ ($H_1$) will result in a net magnetic moment towards $H_1$, resulting in shrinkage of the lattice along the "1" axis but elongation along the "2" axis, resulting in a negative $\lambda_{11}$ and a positive $\lambda_{21}$ with an almost similar magnitude. Only a small amount of out-of-plane moments flopping into the plane can produce small elongation along the "3" axis (thus small $\lambda_{31}$). On the other hand, the field along the "3" axis ($H_3$) can flop all the in-plane moments to "3", resulting in the largest expansion in the plane (thus largest $\lambda_{13}$). All our experimental results are perfectly consistent with this picture. This further suggests that one can tailor the ME properties of the laminate ME media by controlling the magnetic domain structure.

Based on the enhanced $\alpha_{E33}$, we further focused on increasing $\alpha_{E33}$ and related longitudinal ME voltages by optimizing the number of layers and thickness. Figure 3 summarizes the thus obtained maximum values of $\alpha_{E33}$ ($\alpha_{E33,m}$) and the longitudinal ME sensitivity ($\delta V_3/\delta H_{3,m} \equiv nt\alpha_{E33,m}$) as a function of $n$ for both $t = 50$ and 15 μm. The $\alpha_{E33,m}$ values are found to be nearly constant for different $n$, which guarantees almost linearly increasing $\delta V_3/\delta H_3$ with respect to $n$. Therefore, for both thickness values, the interface coupling is maintained with increasing the number of layers up to $n = 25$. Meanwhile, the decrease in the $t$ value from 50 to 15 μm resulted in an enhancement of $\alpha_{E33,m}$ by about a factor of 4. The larger $\alpha_{E33}$ for $t = 15$ μm indicates that the thinner layer is more effective in transferring interface strain into each layer. Based on these MELs with $t = 15$ μm, we could achieve $\alpha_{E33,m} = 15\text{-}18$ mV/cm Oe and $\delta V_3/\delta H_{3,m} = 400 \times 10^{-6}$ V/Oe for $n = 25$. In comparison, the maximum $\alpha_{E33}$ value in the present MEL sample is about twice that of the commercial multilayer capacitor.[22] More importantly, the maximum $\delta V_3/\delta H_3$ is about 60 times larger than that found in ref. 22, which had a parallel capacitor configuration. We note that upon further optimizing thickness and layers, there exists lot of room to improve the $\delta V_3/\delta H_3$.

In conclusion, we have successfully synthesized and measured ME properties of the NFO-BTO magnetoelectric laminates grown by the tape casting method. We showed that the longitudinal ME sensitivity was greatly increased due to the in-plane aligned magnetic moments in the NFO layers and that the ME voltage could also be increased in proportion to the number of layers in a series connected configuration. Our findings offer pathways of engineering ME voltages in mass-produced, environmentally friendly MELs for numerous ME device applications.

This study was supported by NRF through the National Creative Research Initiatives (2010-0018300) and Basic Science Research (2009-0083512) programs and by the Ministry of Knowledge Economy (MKE) through the Fundamental R&D Program for Core Technology of Materials.

**Figure captions:**

**Fig. 1.** (a) Schematic drawing and characteristics of parallel and series connected magnetoelectric laminates (MELs). (b) Structure of the fabricated MEL (top) and its optical microscope image (bottom). The measurement configuration is also shown. (c) X-ray diffraction pattern in a powdered MEL composed of $BaTiO_3$ (BTO) and $NiFe_2O_4$ (NFO) layers. Inset compares the XRD patterns of the MEL before and after grinding into a powder form.

**Fig. 2.** (a) Longitudinal ($\alpha_{E33}$) and transverse ($\alpha_{E31}$) ME voltage coefficients (b) $\lambda_{11}$, $\lambda_{21}$, $\lambda_{31}$, and $\lambda_{13}$. (c) $q_{11}$, $q_{21}$, $q_{31}$, and $q_{13}$, and (d) magnetization vs $H$ curves in the MEL with $n$ layers of BTO and ($n$+1) layers of NFO. Inset of (d) shows the proposed magnetic domain pattern in the NFO layers in our MELs.

**Fig. 3.** Maximum of $\alpha_{E33}$ ($\alpha_{E33,m}$) and $dV_3/dH_3$ ($dV_3/dH_{3,m}$) as a function of $n$ for the ($n$) BTO – ($n$+1) NFO MELs with $t =$ (a) 50 and (b) 15 μm.

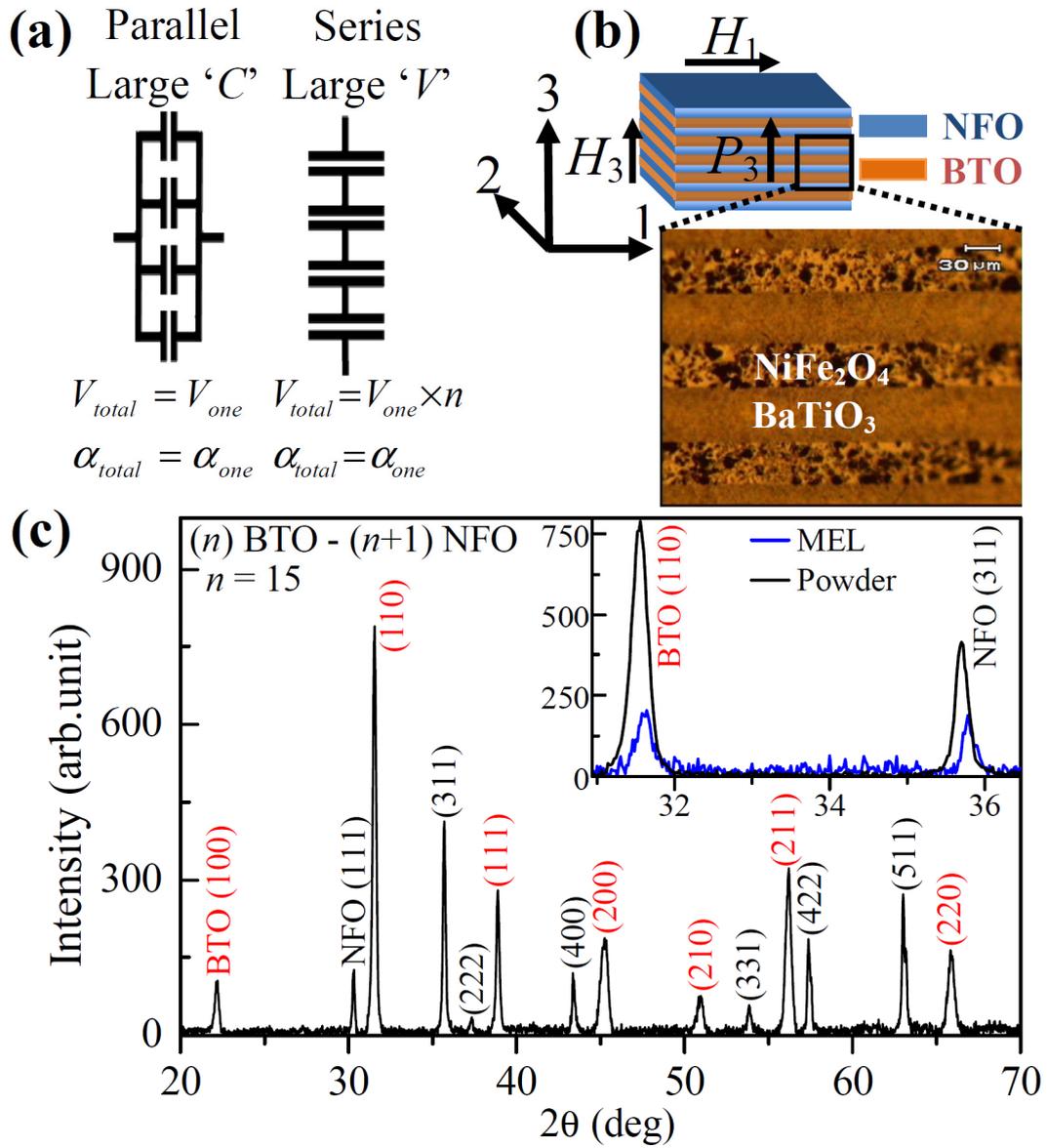

**Fig. 1**

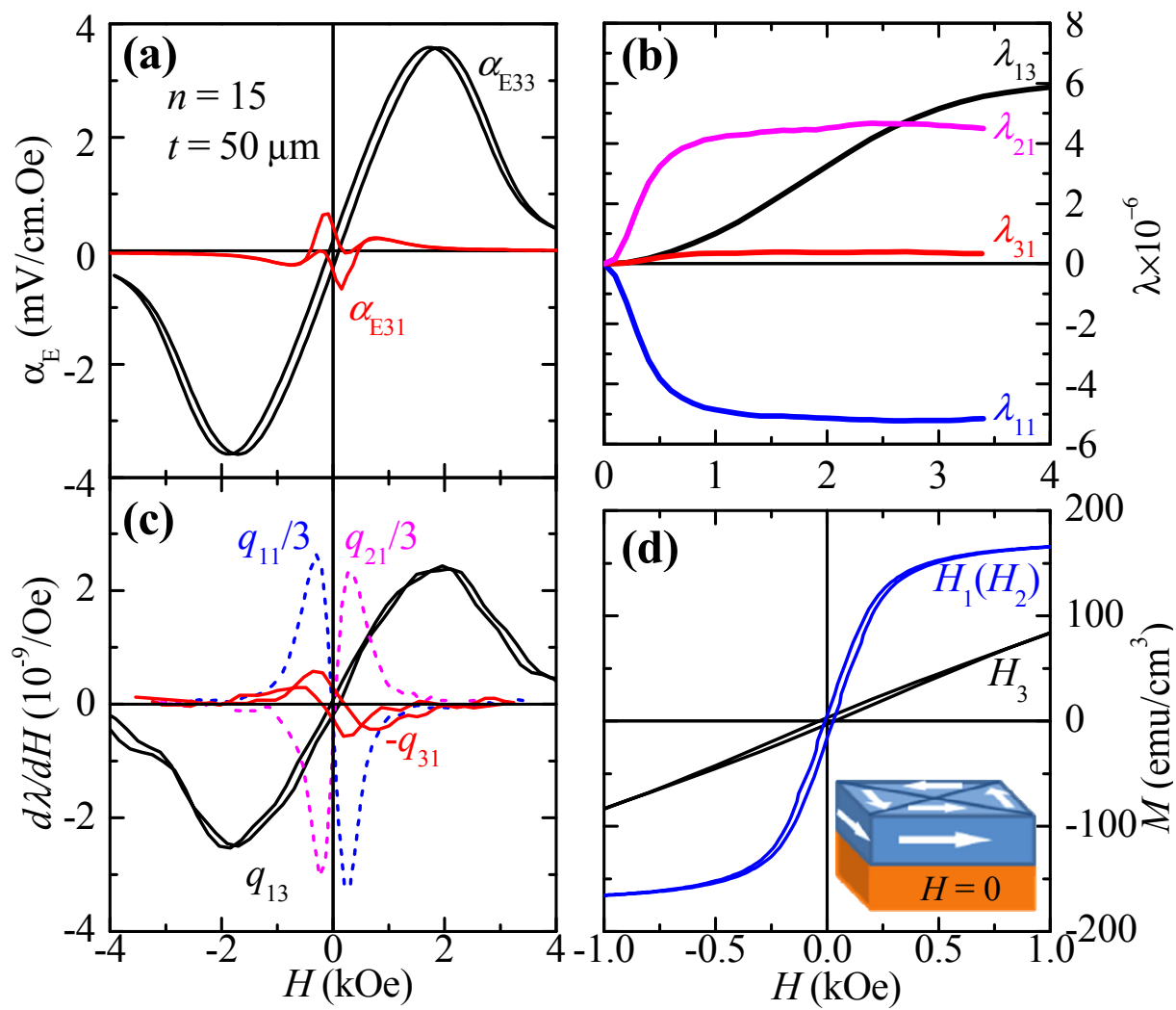

**Fig. 2**

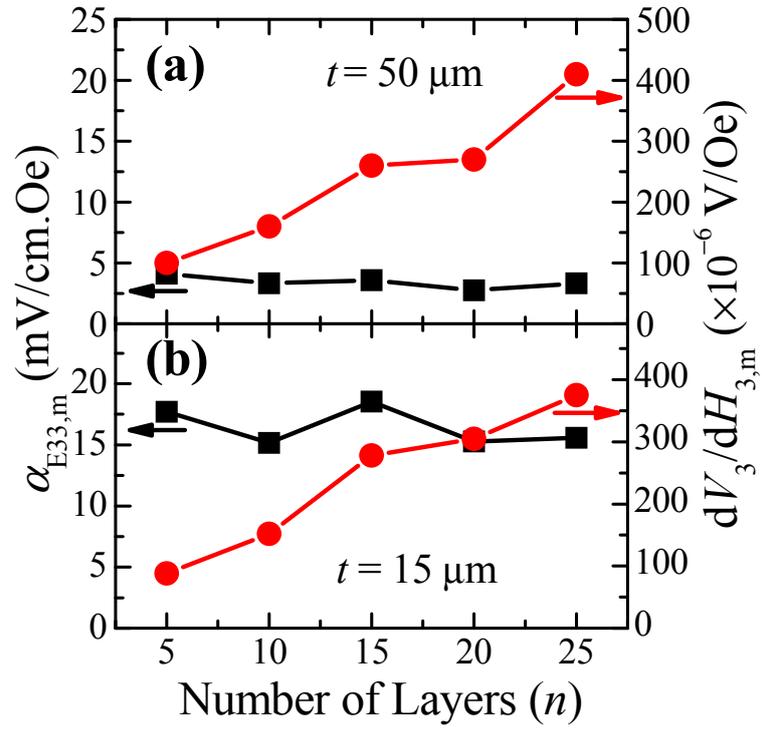

**Fig. 3**